\documentclass[twocolumn]{aa}
\usepackage{graphicx,natbib}
\usepackage{txfonts}
\newcommand{\bq}{\begin{equation}}
\newcommand{\eq}{\end{equation}}

\begin{document}

\title{The blue stragglers formed via mass transfer in old open clusters}

\subtitle{}

\author{B. Tian\inst{1,3}
        \and L. Deng\inst{1} \and Z. Han\inst{2} \and X.B. Zhang\inst{1}}

\offprints{B. Tian}

\institute{National Astronomical Observatories, CAS, Beijing 100012, P.R. China\\
           \email{tianbin@bao.ac.cn}
       \and
           National Astronomical Observatories/Yunnan Observatory, CAS, Kunming, 650011, P.R. China
        \and
       Graduate University of Chinese Academy of Sciences, Beijing, 100049, P.R. China}

  \date{Received ; accepted }

\abstract
  {}
  {In this paper, we present the simulations for the primordial blue stragglers
  in the old open cluster M67 based on detailed modelling of the evolutionary
  processes.   The principal aim is to discuss the contribution
  of mass transfer between the components of close binaries to the blue straggler population in M67.}
  {First, we followed the evolution of a binary of 1.4M$_\odot$+0.9M$_\odot$.
  The synthetic evolutionary track of the binary system revealed that a primordial
  blue straggler had a long lifetime in the observed blue straggler region
  of color-magnitude diagram.
  Second, a grid of models for close binary systems experiencing
  mass exchange were computed from 1Gyr to 6Gyr in order to account for primordial blue-straggler
  formation in a time sequence. Based on such a grid, Monte-Carlo simulations
  were applied for the old open cluster M67.}
  {Adopting appropriate orbital parameters, 4 primordial blue stragglers were predicted
  by our simulations. This was consistent with the observational fact that only a few blue
  stragglers in M67 were binaries with short orbital periods. An upper boundary of the primordial blue stragglers
  in the color-magnitude diagram (CMD) was defined and could be used to distinguish blue stragglers
  that were not formed via mass exchange. Using the grid of binary models, the orbital periods
  of the primordial BSs could be predicted.}
  {Compared with the observations, it is clear that the mechanism discussed in this work alone cannot fully
  predict the blue straggler population in M67. There must be several other processes also involved in the formation of the
observed blue stragglers in M67.}


   \maketitle
%

\section{Introduction}\label{Sect1}
Blue stragglers (BSs) were noticed for the first time in the
globular cluster (GC) M3 \citep{san53}. These ``blue extensions''
lie above the main sequence turn off (MSTO) in the CMD of the GC.
Afterwards, BSs have been observed in stellar systems of various
types and ages. As the brightest members along the main sequence in
the CMD of a given stellar system, BSs should make a considerable
contribution to the blue side of the integrated light of the stellar
system \citep{den99,sch04,xy05}.

\par
Observations show that BSs have higher masses than the stars at the MSTO of the
host clusters \citep{sha97,nem89}. However, the classical stellar evolutionary
scheme shows that single stars are similar in mass to these BSs (more
massive than MSTO) in the clusters should have already evolved away from that
region in the CMD. The unusual characteristics of observed BSs cannot be explained
by the theory of single-star evolution in a coeval stellar system like a
cluster.

\par
Several mechanisms for BS formation have been proposed so far, with all
the established explanations related to dense environments or
binary evolution, namely, direct collisions between stars, mass transfer, or
coalescence in close binary systems \citep{mcc64,str70,hil76}.
Direct physical collision hypothesis was
originally presented by Hills \& Day (1976). They proposed that the
remnant of a collision between two main-sequence stars could produce
a blue straggler. All types of colliding encounters including
single-single, binary-single, and binary-binary have been investigated by
subsequent studies \citep{oue98,lom02,fre04}.

\par
The proposition that BSs could be formed through mass transfer in
close binary systems was presented for the first time by McCrea in 1964.
He suggested that the primary could transfer material to
the secondary through the inner Lagrangian point after becoming a
red giant and filling up its Roche lobe. The material is stripped
off of the envelope of the primary and is accommodated on the
surface of the secondary. Then, as the secondary is gaining material
gradually, it can became a more massive main-sequence star with a
hydrogen-rich envelope. This ``modified'' (or, newly
formed) star can stay on the upper extension of the main sequence
until finishing hydrogen burning in the core.
As a result of mass transfer, the star has its
main-sequence lifetime doubled compared to a normal star with the same
mass. Such a close binary system is called
primordial BS \citep{dav04}. McCrea sketched such a picture and
predicted that the newly formed star could be as bright as about 2.5
magnitudes more luminous than the turn-off point of a cluster. As
shown in the CMDs of star clusters, most BSs do indeed stay
in that region.

\par
The mass transfer between the components of binary systems were
divided into case A, case B, and case C according to the evolutionary
phases when the donor fills up its Roche lobe \citep{kw68}. These
three cases are associated with hydrogen burning in the core, with the
rapid core contraction preceding helium ignition, and with helium
having been ignited in the core.

\par
Chen \& Han (2004) have demonstrated the possibility of BS formation via
mass exchange recently. Their work reveals the evolution of both the
donor and the accretor in a binary system experiencing mass
transfer. They conclude that BSs may be produced in
short orbital-period binaries via either the case A or case B mass-transfer
scheme. They also point out that the abnormalities in the
surface composition can make the accretor bluer.

\par
A large number of BSs in various stellar systems have been observed
\citep{mat91,lan97,van01,san03,ss03}. For example, Mathys (1991)
presents a detailed analysis of eleven BSs in M67. The author points
out that some of the observed BSs in M67 show the characteristics of
a binary system and of some properties of main sequence stars of the
same spectral type, including surface gravities and the abundances
of some chemical elements. These observational properties are
verified by Milone \& Latham \citep{mil92,lat96}. Their long-term
observations confirmed the binary properties of six BSs in M67. Five
out of the 6 BSs are long-period spectroscopic binaries, while the
6th (S1284) is a binary with $P$=4.18 d and $\it e$=0.205. Milone \&
Latham (1992) have suggested that the 6th BS is very likely a
primordial BS candidate. But the eccentricity of S1284 is indeed a
puzzle. The orbit of a short-period binary experiencing mass
exchange should be circularized by tidal effects. Milone \& Latham
(1992) mention that the eccentricity might be an artifact caused by
some sort of line asymmetry in orbital solutions, an accretion disk
formed during the mass transfer, or a third star in the system with
a wide orbit. Among the five long-period binaries, three showed
significant eccentricities, while the other two stars had nearly
circular orbits. Thus Milone \& Latham concluded that there should
be several different mechanisms to form these blue stragglers. When
considering the observations for BSs, Leonard (1996) compared
several hypotheses for the origin of blue stragglers and analyzed
the observational consequences. Following these detailed
comparisons, Leonard also argued that the blue straggler population
in M67 should be produced through several mechanisms.

\par
In general, BSs in a dense stellar environment are
more likely the remnants of direct collisions.
In a sparse stellar environment, however, mass transfer
in primordial binaries (PBs) is believed to be the primary scheme
for BS formation. Ferraro et al. (2004) find that the radial
distribution of the BSs in 47 Tuc appears bimodal, i.e., highly
peaked in the core, decreasing at intermediate radii, and finally
rising again at larger radii. Mapelli et al. (2004) studied the
observed bimodal distribution of BSs in 47 Tuc with a series of
simulations using a dynamical code. They find that the internal
BSs mainly result from collisions and the external ones
are exclusively generated by mass transfer in primordial binaries.
Therefore the explanation of the observed BS distribution of 47 Tuc
requires a combination of primordial BSs and collisional BSs. Piotto
et al. (2004) studied nearly 3000 BSs within 56 GCs, to find that they
show unexpected statistical properties$-$the BS population in a
cluster depends neither on total mass nor on the stellar collision rate of
the host cluster. Davies et al. (2004) analyzed the mechanisms of BS
formation in GCs. They combined the contributions from
both dynamical blue stragglers and primordial blue stragglers. The
results could match the previous observation from Hubble Space
Telescope \citep{pio04}. Thus Davies et al. suggest that the BSs
in GCs should be formed through both collision and mass-transfer
processes.

\par
Collier \& Jenkins (1984) calculated a series of Monte-Carlo
simulations for the binary system evolution of old disc clusters, the
results seemed to favor the formation scheme through mass transfer in
close binaries. With empirical distributions of the initial orbital
parameters of close binary systems, the numbers of BSs produced in
their simulations matched the observed BS number counts well.

\par
Pols \& Marinus (1994) performed Monte-Carlo simulations of close binary
evolution in young open clusters. Their simulations show that the predicted
number of BSs is consistent with the observed value of the clusters younger
than about 300 Myr. However, the simulations of the clusters with ages between
300 and 1500 Myr cannot produce enough BSs to fit the observations of
the clusters with corresponding ages. Thus Pols \& Marinus suggest that there
must be other formation processes functioning at these ages.

\par
Hurley et al. (2001) applied their N-body code to model the BS
population in M67. Their code includes the cluster dynamics besides
modeling the stellar and binary evolution. Compared with binary
population synthesis, the N-body simulation can produce twice the number
of BSs, which agrees with the observations. Thus Hurley et
al. (2001) argue that the dynamical cluster environment of M67
plays an important role in producing a consistent number of BSs.
The authors also show that the BSs predicted by this simulation of M67
are due to several formation processes. Later, Hurley et al. (2005)
present a dynamical simulation using a direct N-body model for M67.
At 4Gyr, 20 BSs are obtained, 11 of which are considered as
direct mergers of two MS stars in primordial binaries.
Nine of these cases were formed from Case A mass transfer, while
the other two were collisions in eccentric binaries.
The remaining 9 BSs were in binaries. Only two of these are primordial
binaries. In their model, only 7 of 20 BSs formed from PBs were not affected
by the cluster environment. Thus they
suggest that the formation of the BS population in M67 is dominated
by both mass transfer and cluster dynamics.

\par
In this paper, a detailed modelling of primordial BSs is
presented. Such a treatment shows some advantages over the previous
population synthesis schemes when considering non-equilibrium
stars. A Monte-Carlo simulation based on these models was carried out to
study the number of BSs in old open clusters from the primordial
channel. In Sect. 2, we describe the evolution of primordial BSs.
The results of our example model (1.4M$_\odot$+0.9M$_\odot$) are
presented. In Sect. 3, the results of Monte-Carlo simulations
are presented. A summary and conclusions of our work are given in the final section.

\section{The model of primordial blue stragglers}
We adopt Eggleton's (1971, 1972, 1973) stellar evolution code to
simulate the primordial BS evolution in this work. The code has been
updated in subsequent work \citep{han94,pol95,pol98}
and is described by Han et al.(2000). The code uses a
self-adaptive non-Lagrangian mesh and adopts Opacity Library (OPAL)
radiative \citep{igl96} and molecular opacities \citep{alx94}.
Convective overshooting hardly affects
the evolution of low-mass stars \citep{pol98}, so it is not considered in this
work. The mass and angular momentum (AM) of the binary systems are
assumed to be conservative in our calculations.

\par
Roche lobe overflow (RLOF) is included as a boundary condition in
Eggleton's code. When RLOF takes place, the mass transfer rate is
calculated via the relation

\bq
\frac{{\rm{d}}m}{{\rm{d}}t}=C\times\max{[0,(\frac{R_{\rm{S}}}{R_{\rm{L}}}-1)^3]}
\eq
where $R_{\rm{S}}$ is the radius of the donor, and $R_{\rm{L}}$ that
of the corresponding Roche lobe. In this work, we take
$C=$500M$_{\odot}$yr$^{-1}$ to keep a steady RLOF. As discussed by
Han et al. (2000), the donor overfills its Roche lobe as necessary
but never overfills its lobe by much:
$(\frac{R_{\rm{S}}}{R_{\rm{L}}}-1)\leq0.001$. To avoid complications
due to uncertainties in the contact systems, only the stable RLOFs are
considered in this work. Our calculations are terminated
once merger occurs, but the information up to the stage of contact
is kept in our grids. Subsequent evolution is very important, so
further study is needed.

\par
We have neglected the influences due to tidal evolution, magnetic braking, and stellar spins
on the evolution of binary systems for the following reasons.
For eccentric binaries, the modelling of tidal circularization
will be crucial if mass transfer happens. However, compared with adopting
eccentric orbits and considering the tidal effects, adopting initially circular
orbits will not affect the outcomes of a large binary population synthesis.
Combining the ongoing processes of magnetic braking and
tidal synchronisation will affect the mass transfer rate and merger
timescale \citep{ste95}. For simplicity, the present model does not include
treatment of these effects, but they will be taken up in a subsequent work.

\par
When calculating the evolution of a donor, mass transfer rates at
the corresponding ages are recorded and then used as input for
adjusting the mass of the accretor, such that the evolution of the
two components are synchronized. The accreted matter is assumed to
be deposited onto the surface of the secondary with zero falling
velocity and distributed homogeneously over the outer layers, and
the evolution of the components is followed as usual for stellar
evolution. Previous population synthesis studies have adopted much
more simplified stellar evolution schemes. These rapid codes use
analytic formulae to approximate the main characters of the stellar
evolution. With them, the information on stars, such as radius,
luminosity, and age, can be estimated empirically at high efficiency
but at the cost of losing physical details and accuracy. Our
calculations are based on realtime stellar evolution calculations.
We can provide more details about the components of the binary
systems, such as the information about nucleosynthesis in the core
and the chemical abundance of stars.

\par
We now present a system of 1.4M$_\odot$+0.9M$_\odot$ as an
example. The initial orbital separation is 5.0R$_\odot$ and the solar
composition $[Z=0.020,Y=0.280]$ is adopted. A series of previous
studies \citep{zah66,tas88,gol91} dealt with the orbital circularization
of close binaries caused by tidal interaction between the two
components. It has been proved that the tidal interaction can
circularize the short-period binaries, meaning that the orbital
eccentricity of a close binary will decay during the course of
the dynamical evolution of the system. For simplicity, we assume the
orbital eccentricity $\it e$$=0$ at zero age main sequence (ZAMS) for all cases considered in
this work.

\par
\begin{figure*}
\hspace{4cm}
\includegraphics[angle=-90,width=70mm]{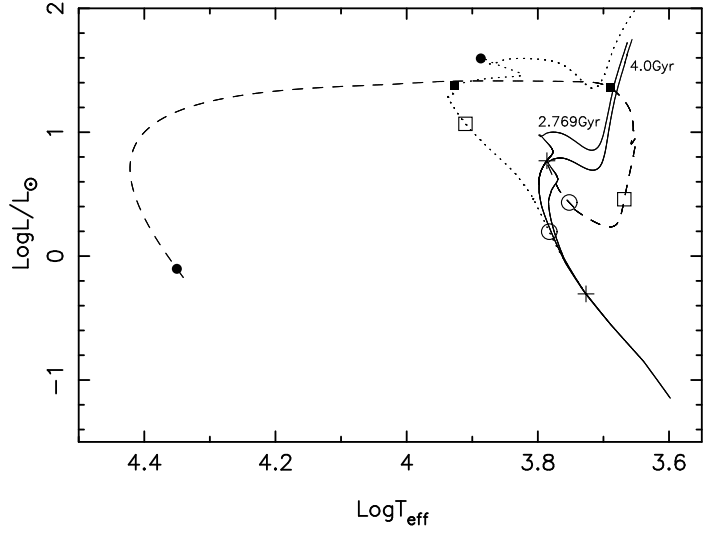}
\caption{The evolutionary tracks of the two components of our
example binary after mass transfer begins. The dashed and dotted
lines are the evolutionary tracks of the donor and the accretor.
On both tracks, the crosses, open circles, open squares, filled
squares, and filled circles show the positions in order: mass
exchange begins, the mass ratio of the system equals 1, the system is
4.0Gyr old (the age when M67 is observed), mass transfer
terminates, and the core H-burning of the secondary ends. The two thin
solid lines are the 2.769Gyr and 4.0Gyr theoretical isochrones.}
\label{fig:tracks}
\end{figure*}

The evolutionary tracks of the two components
(1.4M$_\odot$+0.9M$_\odot$) of the example system beginning with the onset
of mass transfer are given in Fig.~\ref{fig:tracks}.
The two components of the
example binary depart from the regular evolutionary tracks of
single stars after mass transfer occurs at an age of 2.769Gyr.
Because the binary system begins mass transfer at 2.769Gyr and the
age of M67 is 4.0Gyr, the 2.769Gyr and 4.0Gyr theoretical
isochrones are also plotted as a reference with solid lines in
Fig.~\ref{fig:tracks}. When RLOF begins, the donor is still a main
sequence star, so that our example binary follows the case A mass
transfer scheme. The donor reaches the bottom of the giant branch
after Hydrogen in its core becomes exhausted. Then the donor
evolves along the giant branch. Eventually, the donor becomes a
white dwarf. Due to mass accretion, the luminosity of the
accretor increases. After mass transfer terminates at 4.423Gyr,
the accretor follows the evolutionary behavior of a single star
again with the corresponding mass. Table~\ref{tab:results}
presents the main parameters of the binary for six key epochs
in order: ZAMS, mass exchange begins, the mass ratio of the system equals 1,
the system is 4.0Gyr old (the age when M67 is observed), mass transfer
terminates, and the core H-burning of the secondary ends.

\begin{table*}
\caption{Main results of the example binary (1.4M$_\odot$+0.9M$_\odot$)}
\label{tab:results}
\begin{center}
\begin{tabular}{cccccccccc}
\hline
Epoch & Age & $P$ & $a$ & Mass & $\lg{(L/L_\odot)}$ & $\lg$$T_{\rm{eff}}$ & $X_{\rm{C}}$ & $Y_{\rm{C}}$ & $\dot{M}$\\
      & (10$^9$yrs) & (d) & (R$_\odot)$ & (M$_\odot)$ & & &  &    & (M$_\odot$yr$^{-1}$)\\
\hline
 1&  0.000 & 0.8543 & 5.0000  & 1.4000 & 0.4984 & 3.8067  & 0.700 & 0.280  & 0.0\\
  &        &        &         & 0.9000 & -0.3802 & 3.7165 & 0.700 & 0.280 & 0.0\\
 2&  2.769 & 0.8543 & 5.0000  & 1.4000 & 0.7307 & 3.7845 & 0.193 & 0.788  & 0.0\\
  &        &        &         & 0.9000 & -0.3048 & 3.7266 & 0.574 & 0.406 & 0.0\\
 3&  3.209 & 0.7388 & 4.5386  & 1.1500 & 0.4328 & 3.7521 & 0.096 & 0.884  & $1.62\times{10^{-10}}$\\
  &        &        &         & 1.1500 & 0.1968 & 3.7826 & 0.532 & 0.448  & $1.62\times{10^{-10}}$\\
 4&  4.000 & 2.1938 & 9.3761  & 0.5157 & 0.4589 & 3.6682 & 0.000 & 0.980  & $1.16\times{10^{-9}}$\\
  &        &        &         & 1.7843 & 1.0680 & 3.9104 & 0.562 & 0.417  & $1.16\times{10^{-9}}$\\
 5&  4.423 & 12.968 & 30.653  & 0.2480 & 1.3611 & 3.6891 & 0.000 & 0.000  & 0.0\\
  &        &         &        & 2.0520 & 1.3781 & 3.9274 & 0.4000 & 0.580  & 0.0\\
 6&  4.864 & 12.968 & 30.653  & 0.2480 & -0.2128 & 4.3353 & 0.000 & 0.000  & 0.0\\
  &        &        &         & 2.0520 & 1.5954 & 3.8869  & 0.000 & 0.980  & 0.0\\
\hline
\end{tabular}

\medskip
The columns are (1) the model serial number, (2) the age, (3) the
period of the binary system, (4) the separation between the two
components, (5) the masses, (6) the luminosities, (7) the effective
temperatures, (8) hydrogen abundance in the core, (9) helium
abundance in the cores, (10) mass transfer rate
\end{center}
\end{table*}

\par
In most cases, binary systems cannot be resolved visually in observations, so
that we cannot put the components in a CMD separately. As such, we need to
synthesize the total light of the system including contributions from both
components to get the evolutionary track of the binary system as a whole in the
CMD. In our theoretical model, we obtained the intrinsic parameters,
including effective temperature, luminosity, mass, and surface gravity of the
two components along their respective evolutionary tracks. Given effective
temperature and surface gravity, and assuming solar abundance, we can obtain
theoretical spectra \citep{lej97,lej98} of two components along their
evolutionary tracks. By convolving filter response and the corresponding
spectra, the composite track of the system in the CMD can be derived as

\bq m=-2.5\lg\int^{\lambda_2}_{\lambda_1}
F({\lambda})\cdot\varphi({\lambda})\rm{d}{\lambda}+C
\eq
Where $F({\lambda})=F_1({\lambda})+F_2({\lambda})$ is the total flux
of the two components, $\varphi({\lambda})$ is the response function
of a filter, ${\lambda_1}$ and ${\lambda_2}$ give the range of response
$\varphi({\lambda})$, $C$ is a constant and can be determined by
calibrating with observation of the Sun.

\par
\begin{figure*}
\hspace{4cm}
\includegraphics[angle=-90,width=70mm]{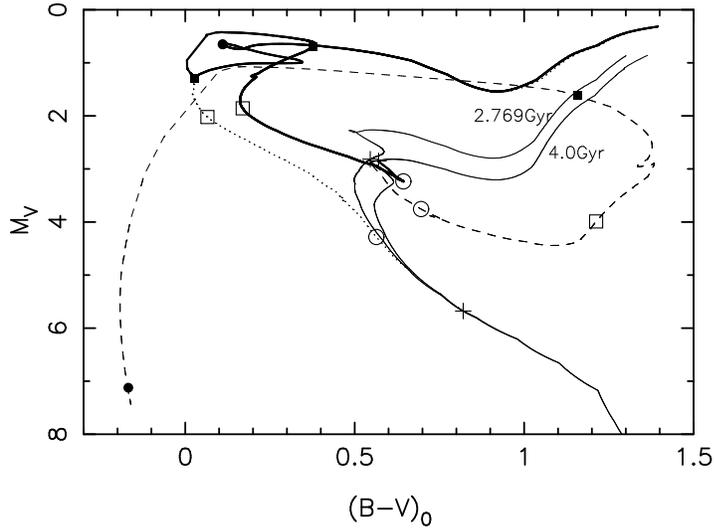}
\caption{The synthetic evolutionary track of the example binary
system in the CMD. The thick solid line is the synthetic evolutionary
track of the binary. The dashed and dotted lines are the
evolutionary tracks of the donor and the accretor. The two thin
solid lines are the 2.769Gyr and 4.0Gyr theoretical isochrones. All
the other symbols have the same meanings as in Fig.~\ref{fig:tracks}.}
\label{fig:cmd}
\end{figure*}
\begin{figure*}
\hspace{4cm}
\includegraphics[angle=-90,width=70mm]{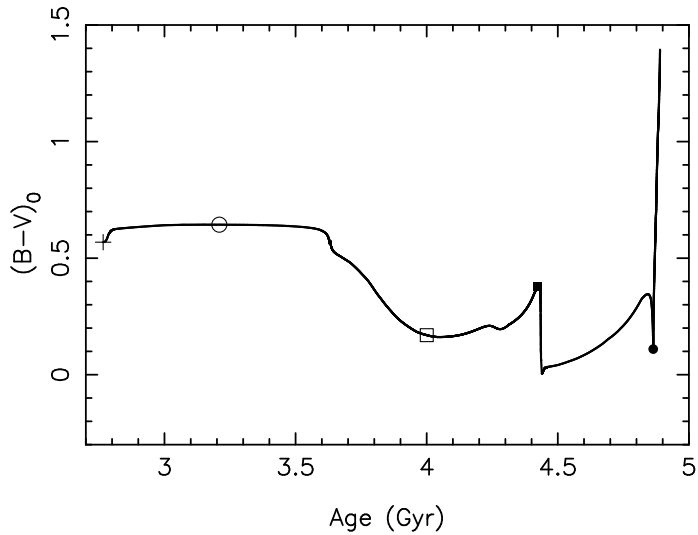}
\caption{Evolution of the synthetic color of the example binary
system. All the symbols have the same meanings as in
Fig.~\ref{fig:tracks}.} \label{fig:color}
\end{figure*}

Using the response functions of the Johnson $B$ and $V$ filters, we
can get the Johnson $B$ and $V$ magnitudes of the binary system. In
Fig.~\ref{fig:cmd}, the synthetic evolutionary track of the binary
system in CMD is shown. Before the mass ratio reverses, the higher
luminosity of the donor dominates the synthetic evolutionary track
of the binary system. After the mass ratio reverses, the accretor
becomes dominant in luminosity and color. After the mass ratio
becomes 1, the binary system evolves towards the blue straggler
region of the CMD. In the subsequent evolution, the binary system
spends 1.228Gyrs in the region that is bluer than MSTO of the 4.0Gyr
isochrone until the secondary leaves the main sequence. Thus the
binary system has enough lifetime in that color remaining to be
observed as a BS. Evolution of the synthetic color of the example
binary system is given in Fig.~\ref{fig:color}. Obviously, most of
the lifetime of the binary after mass transfer starts is spent on
the blue side of the CMD (defined as (B-V) $<$ 0.545 -- the color of
MSTO of the 4.0Gyr theoretical isochrone).

\par
To simulate the population of the primordial BSs in old open
clusters, a grid of the models of close binaries experiencing mass
exchange is needed. To this aim, we calculated various combinations of binary
parameters that can have mass exchange triggered for the donor
masses of 0.1M$_\odot$ to 2.0M$_\odot$. The mass
range of the accretor is from 0.1M$_\odot$ to the mass of the
corresponding donor. The mass intervals of the donors and the
accretors are both 0.1M$_\odot$. To cover all possible cases of
close binaries, the orbit separation ranges from 1.0 to
50.0R$_\odot$ with a grid interval of 1.0R$_\odot$. Following the
numerical scheme described above, a grid for six ages from 1.0Gyr
to 6.0Gyr is obtained in Table 2 (available in its electronic form).
Given the position in CMD of
a primordial BS, one can estimate the parameters of the components
of the binary systems with this table.

\section{Monte-Carlo simulations of the primordial BSs in M67}
\subsection{Initial parameters of the cluster M67}
With the grid of primordial BS models, it is possible to statistically study the
total number and distribution of primordial BSs in CMD
for a given cluster. Here, we study the BSs in the
old open cluster M67 using the Monte-Carlo method. Sandquist's (2004)
estimation of the age of M67 is adopted. As claimed by Sandquist,
the main features of the cluster in the CMD up to the subgiant
branch can be reproduced well by a 4-Gyr isochrone, with an
uncertainty lower than 0.5 Gyr. The metallicity of M67 is believed
to be solar \citep{hob91}, while Hurley et al. (2005) point out that
the current total mass of M67 is about 1400M$_\odot$. By
considering the photometric limits in the previous work, the
actual mass of M67 ought to be slightly higher. Owing to so much
detailed work on the basic parameters of the cluster and on the
individual objects in the cluster including its BS population,
models made to BSs in this cluster can be constrained well.
We concentrate on the formation of primordial BSs in the cluster
and present numerical simulations for its type.
As discussed in Sect. 2, the orbital eccentricities
of close binaries in our calculations are all assumed to be $\it e=0$.
The distributions of the other initial parameters of binary
systems are listed as follows:

\begin{enumerate}

\item Hurley et al. (2005) suggest that a system including 12000
single stars and 12000 primordial binaries at its birth can
reproduce the observed parameters of M67 in the present. Such a
configuration is adopted in this work. These numbers include the
stars that are gravitationally evaporated in the earlier history
of M67. Since we are considering primordial binaries that may
eventually form BSs, such binary systems are presumably massive
points in the cluster and therefore are unlikely to be affected by
dynamical evaporation. Adopting the initial number of binaries in
our case should not be very different from models including
dynamical evolution. This issue will be discussed in
Sect~\ref{sec:results}.

\item The initial mass function (IMF) of the binary systems in a
cluster has been discussed by Kroupa, Tout \& Gilmore (1991,
hereafter KTG). According to the KTG IMF, the initial masses of the
binary systems can be given by the generating function
 \bq
M(X)=0.33[\frac{1}{(1-X)^{0.75}+0.04(1-X)^{0.25}}-\frac{1}{1.04}(1-X)^2]
\eq

where $M(X)$ is the binary mass in units of M$_\odot$, and X is a
random number with a uniform distribution between 0 and 1. The
original single-star population in M67 is assumed to have a mass
coverage from 0.1M$_{\odot}$ to 50.0M$_{\odot}$. Thus the binary
mass $M(X)$ is constrained by the limits of 0.2M$_{\odot}$ and
100.0M$_{\odot}$.

\item The uniform distribution of mass ratio assumed by Hurley et
al. (2001) is adopted. The mass ratio is between
$\max[0.1/(M(X)-0.1),0.02(M(X)-50.0)]$ and 1.

\item The flat distribution of orbital separations from Pols \&
Marinus (1994), $\Gamma(a)\propto{a^{-1}}$, is used here. We assume
that the lower limit of the orbital separations is defined by the
minimum size of the Roche lobe when RLOF takes place at ZAMS. An upper
limit of 50 au as given by Hurley et al. (2005) is adopted.

\end{enumerate}

\subsection{Results}
\label{sec:results} Given the initial input parameters for the
cluster, KTG IMF, mass ratio, and orbit separation of the
binary population at the initial stage, Monte-Carlo simulations
for the M67 BS population can be carried out. We adopt linear
interpolation for the magnitudes and $B-V$ colors of all the
binaries at 4.0Gyr using the grid listed in Table 2. The first
interpolation parameter is orbital separation, the second
mass ratio, and the last the mass of the donor. If the
initial parameters of the binaries are not in the range of the
grid, we assume that mass exchange does not occur in these
binaries.

\par
\begin{figure*}
\hspace{4cm}
\includegraphics[angle=-90,width=70mm]{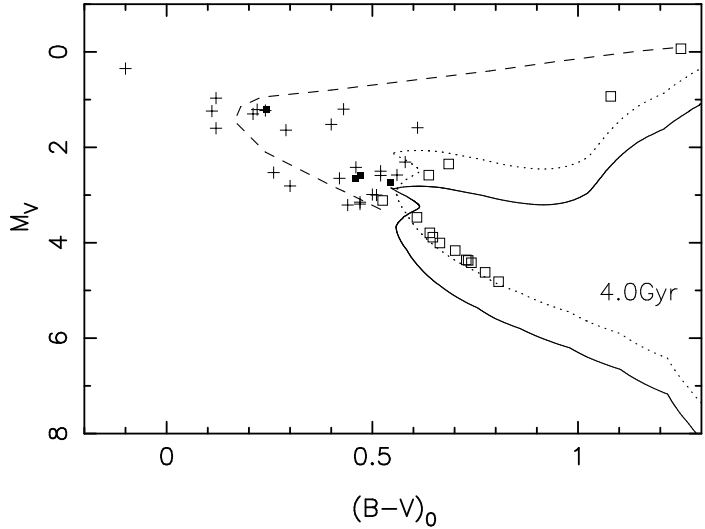}
\caption{The simulated CMD for the primordial BSs in M67. The solid
line is the 4.0Gyr theoretical isochrone. The dotted line is the
equal mass photometric binary sequence. The filled squares are 4
primordial BSs. The open squares are the other 15 PBs
experiencing mass exchange. The crosses are the observed BSs from
Deng et al. (1999). The dashed line shows the upper edge of the
simulated primordial BSs region.}
\label{fig:M67}
\end{figure*}

In Fig.~\ref{fig:M67} we present the results of the Monte-Carlo
simulations for M67. In our simulations, we got 19 PBs
experiencing mass exchange. There are 4 in the region of BS with
respect to the 4Gyr isochrone among these PBs, which can be
identified as primordial BSs. There are 4 primordial BSs shown
in Fig.~\ref{fig:M67}, and another 15 PBs. It also shows the observed BS
sample in this cluster from Deng et al. (1999). The parameters of
the 19 mass transfer PBs are listed in Table~\ref{tab:M67}.
According to the grid, the formation process of the mass transfer PBs
is continuous
during the passive evolution of the stellar population in a cluster.
The binaries listed in Table~\ref{tab:M67} at 4 Gyr
were produced via mass exchange in its earlier history. The
15 PBs plotted in Fig.~\ref{fig:M67}
could be either progenitors or descendants of the primordial BSs.
That means some of these PBs, whose mass transfer is going
on and where the secondary has not yet got enough matter, might become
primordial BSs later, and some were primordial BSs in an earlier
epoch and have already left the region of primordial BSs in CMD at 4
Gyr. Although our model does not evolve dynamically, the BSs
generated in our simulations are mostly from massive binary systems.
Considering that there is a mass segregation effect due to the tidal force of the
Galaxy, the stars lost to dynamical evaporation are mostly low
mass points in the cluster, while the massive ones, however, tend to sink
into the center of the cluster and will be conserved as the cluster
ages. This is consistent with observational results
\citep{mat86,fan96}. The progenitors of the primordial BSs
are very likely the most massive stars. Although dynamical
evolution, such as exchange interactions, can affect the evolution
of some massive binaries, we would like to argue that the number of
BSs predicted in our simulations primarily depends on the input
number of the primordial binaries at zero age.

\setcounter{table}{2}
\begin{table*}
\caption{The parameters of the PBs from our Monte-Carlo simulations}
\label{tab:M67}
\begin{center}
\begin{tabular}{cccccccc}
\hline
$M_1$ &  $q$ & $a$ & $B$ & $V$ & $B-V$ & Comments\\
(M$_\odot$) &  $(M_2/M_1)$ & (R$_\odot$) & & & & \\
\hline
1.4054 &  0.8058 &  4.89 &  1.4554 &  1.2137 &  0.2417 &  PBS  \\
1.4775 &  0.6083 &  4.08 &  3.2915 &  2.7474 &  0.5441 &  PBS  \\
1.4038 &  0.5363 &  4.78 &  3.1078 &  2.6499 &  0.4579 &  PBS  \\
1.4713 &  0.3529 &  5.15 &  3.0676 &  2.5974 &  0.4702 &  PBS  \\
       &         &       &         &         &         &      \\
1.5008 &  0.7213 &  5.20 &  1.1844 & -0.0657 &  1.2501 &  DPBS \\
1.4668 &  0.7041 &  5.18 &  2.0120 &  0.9324 &  1.0796 &  DPBS \\
       &         &       &         &         &         &      \\
1.5912 &  0.4175 &  5.23 &  3.2210 &  2.5838 &  0.6372 &  PPBS or PB    \\
1.1887 &  0.6201 &  2.61 &  5.0869 &  4.3587 &  0.7283 &  PPBS or PB    \\
1.3033 &  0.9402 &  4.41 &  3.0353 &  2.3497 &  0.6856 &  PPBS or PB    \\
1.1998 &  0.6945 &  2.88 &  4.6690 &  4.0048 &  0.6642 &  PPBS or PB    \\
1.1152 &  0.9761 &  3.16 &  4.0771 &  3.4686 &  0.6085 &  PPBS or PB    \\
1.1954 &  0.6037 &  2.94 &  5.1030 &  4.3699 &  0.7330 &  PPBS or PB    \\
1.3748 &  0.5837 &  4.18 &  3.6392 &  3.1142 &  0.5250 &  PPBS or PB    \\
1.1952 &  0.5910 &  2.63 &  5.1598 &  4.4190 &  0.7408 &  PPBS or PB    \\
1.1818 &  0.5404 &  3.44 &  5.3953 &  4.6211 &  0.7742 &  PPBS or PB    \\
1.2600 &  0.5698 &  3.01 &  4.8657 &  4.1641 &  0.7016 &  PPBS or PB    \\
1.1142 &  0.8633 &  3.27 &  4.5303 &  3.8832 &  0.6471 &  PPBS or PB    \\
1.3256 &  0.5125 &  3.87 &  4.4328 &  3.7931 &  0.6397 &  PPBS or PB    \\
1.2031 &  0.4519 &  3.35 &  5.6236 &  4.8168 &  0.8069 &  PPBS or PB    \\
\hline
\end{tabular}

\medskip
The columns are : (1) the initial mass of the donor, (2) the initial
mass-ratio of the binary, (3) the initial orbital separation between
two components, (4) the Johnson $B$ magnitude, (5) the Johnson $V$
magnitude, (6) the $B-V$ color, (7) the comment on PB property:
PBS--primordial blue straggler, DPBS--descendant of primordial blue
straggler, PPBS--progenitor of primordial blue straggler,
PB--photometric binary

\end{center}

\end{table*}

\par
\begin{figure*}
\hspace{4cm}
\includegraphics[angle=-90,width=70mm]{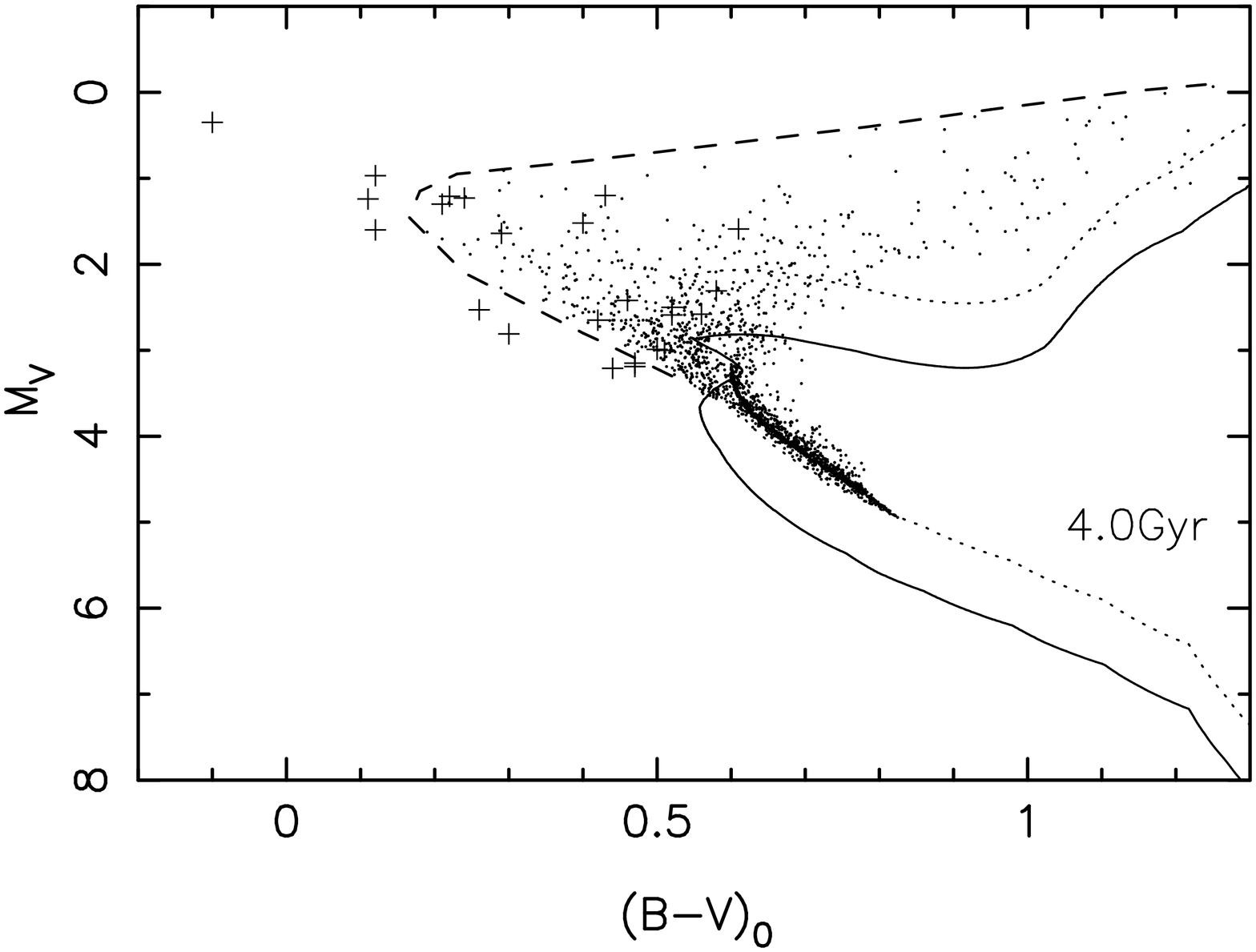}
\caption{The simulated CMD for the primordial BSs in a stellar
system with enhanced primordial binaries. The points are the
PBs experiencing mass exchange. The solid line is the 4.0Gyr
theoretical isochrone. The dotted line is the equal mass photometric
binary sequence. The crosses are the 24 photometrically selected BSs
\citep{den99}. The dashed line is the upper edge of the primordial
BSs region.}
\label{fig:enhanced}
\end{figure*}

In the grid of BS models built with mass exchanging binaries, we
found that there seems to be an upper limit for both coordinates of
the CMD for a given age. In order to investigate such an upper
boundary, we followed the same simulation method as above, but
augmenting the total number of primordial binaries, thereby raising
the total number of binaries that enter BS region in the case of
M67. The simulationin in which the number of primodial binaries has
been raised $(N=1200000)$ is shown in Fig.~\ref{fig:enhanced}, and
1984 PBs experiencing mass transfer were obtained. By counting the
number of the binaries staying in BS region, we find about $11\%$
primordial BSs among 1984 PBs. Most of these PBs stay in between the
turn off and about 2 magnitudes below the turn off of the 4.0Gyr
theoretical isochrone. In our simulations for M67, there are 11 PBs
remaining in that region. The observed CMD \citep{fan96} shows that
quite a few stars still remain in such a region, the characteristics
of these stars are still uncertain, most of them should be
photometric binaries, and some of these stars could be the low mass
primordial binaries that also experienced mass exchange.

\par
An empirical upper limit for the primordial BSs can be drawn from the
simulation(Fig.~\ref{fig:enhanced}). This upper
limit in both luminosity and temperature are defined by the upper
profile of the results, also visible in
Fig.~\ref{fig:M67}. We can define a primordial BS
region that is confined between this upper limit and the given
isochrone. In other old open clusters, we can also define such a
primordial BS region. The observed short period binaries staying in
the region should be primordial BSs. According to our
grid of the models of those close binaries experiencing mass exchange, the blue edge of
the primordial BS region shifts red-wards following the evolution of
a cluster. The blue edge shifts very slowly near $(B-V)=0.1$ from
1.0Gyr to 4.0Gyr, and reaches the bluest color of $(B-V)=0.03$ at
4.6Gyr. The data of the binary systems at 4.6Gyr
are from our detailed modelling of the primordial BSs.
After 4.6Gyr, the blue edge moves to the red side of the CMD
and reaches the position $(B-V)=0.36$ at 6.0Gyr.

\par
A sample of 24 photometrically selected BSs \citep{den99} are
plotted in Fig.~\ref{fig:enhanced}, and their parameters
are listed in Table~\ref{tab:parameters}. The
orbital parameters of 7 stars have already been measured
\citep{mil92,lat96,san03}, and S1284 has been shown to have a short
orbital period ($P$=4.18 d). Chen \& Han (2004) proved that S1284
could have been formed via mass exchange. But they also point out that
the non-zero eccentricity of S1284 is a puzzle, because tidal
effects would circularize the orbit of a binary with a short
orbital period in the mass transfer scheme. This problem was discussed
by Milone \& Latham (1992) who mention that the eccentricity
might not be real, might be affected by an accretion disc, or a
distant third star in a wide orbit. Sandquist et al. (2003) point
out that S1082 should be a triple system made up of a close binary
($P$=1.068 d) and another star. The eccentricity of S1082 in
Table~\ref{tab:parameters} is for the orbit of the third star. The
corresponding period is 1189d, which is too long to have any
important dynamical effects on the close binary system. The primary
of the close binary in S1082 is a BS. S1284 and the close binary in
S1082 are the only BSs in close binary systems that can be accounted
for by mass exchange mechanism.

\par
\begin{figure*}
\hspace{4cm}
\includegraphics[angle=-90,width=70mm]{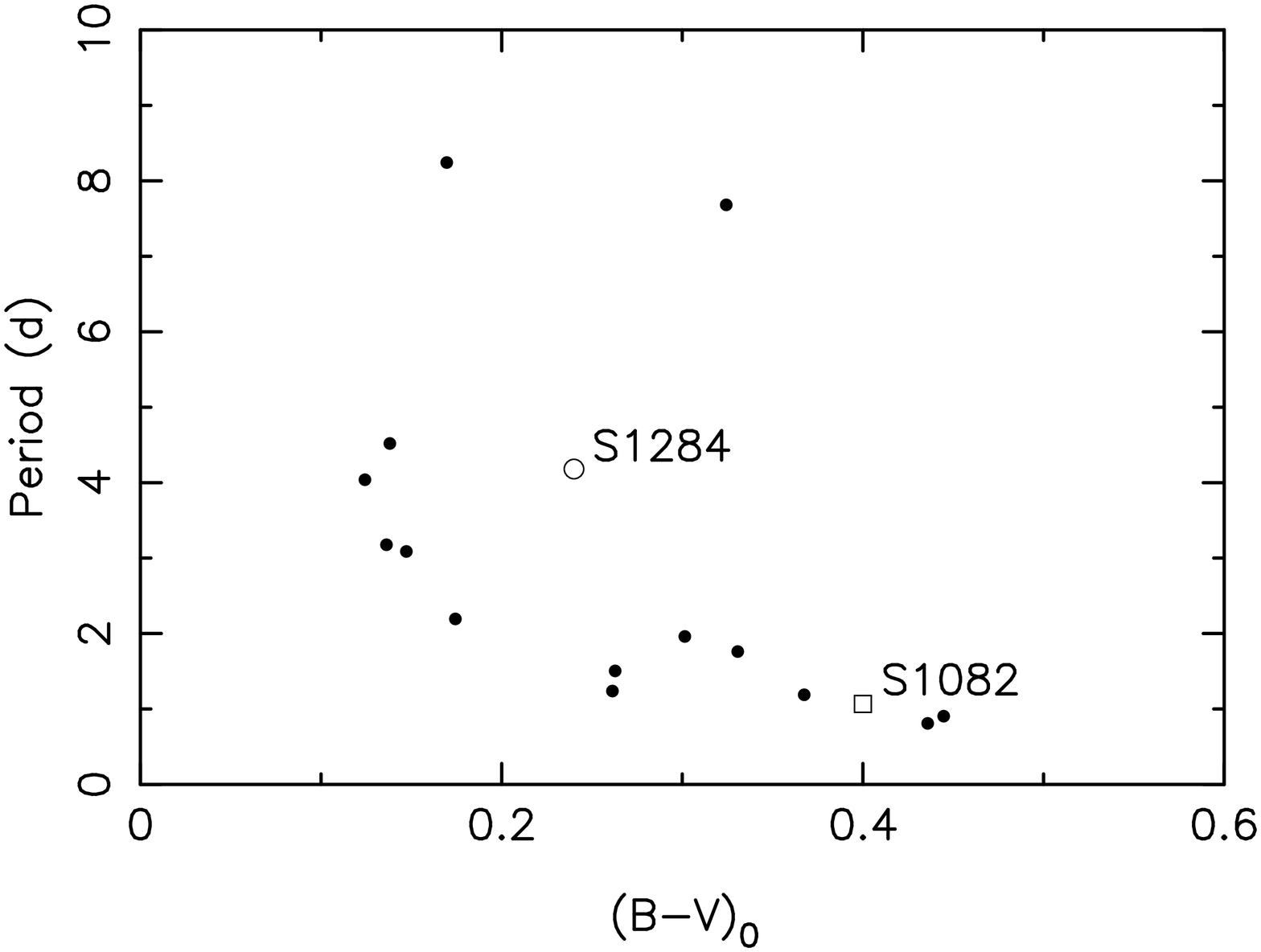}
\caption{The period distribution of the grid of primordial BSs at 4.0Gyr.
The open circle and square show the positions of S1284 and S1082.}
\label{fig:period}
\end{figure*}

The period distribution of the grid of primordial BSs at 4.0Gyr is
given in Fig.~\ref{fig:period}. The grid of the binaries seems to
have two separate branches, one below 5 days and the other over 7
days. The first branch shows an obvious trend; i.e. the bluer the
color of the system, the longer the period. The other one is likely
to behave similarly but have a different slope and some offsets in
period. As the grid is not fine enough for the second branch in this
work, no details can be given at present. Work on this point is in
progress and will be presented in a future paper. However, the 2
observed primordial BSs with short periods S1284 and S1082 are
already well confined in the current grid, as clearly shown in
Fig.~\ref{fig:period}, where the periods of the two observed BSs
match the trend of the first branch closely.

\par
Most of the observed BSs in binary systems show high eccentricity.
This contradicts the assumption $(e=0)$ in our model. But we
cannot confirm that this high eccentricity is the initial nature of these
binaries. Hurley et al. (2005) suggest that dynamical encounters within the
cluster environment and perturbations from nearby stars or binaries could alter
the orbit parameters of the primordial binaries.
Nonetheless, we point out that some of these BSs with high eccentricity
might have been created via wind accretion in wide binaries. In that case,
the orbits can be wide enough to remain eccentric.

\par
There are 9 observed BSs that stay above the upper limit defined above.
We tend to conclude
that these stars should not be the primordial BSs, so that the true nature of these BSs
should be understood by other formation mechanisms. However, as the blue and
upper boundaries of the primordial BS region changes with the age of a cluster,
our conclusion that the BSs are above the limit is not exclusive. For instance, the 9
stars would be covered by the primordial BS region, if the real age of M67 were
around 4.5Gyr. Considering the distribution of orbital inclinations and the
current number of BSs in short period binary systems, the ratio of primordial
BSs in M67 is low, which agrees with the predicted number of BSs in our
simulations.

\setcounter{table}{3}

\begin{table*}
\caption{The parameters of 24 observed BSs in M67.}
\label{tab:parameters}
\begin{center}
\begin{tabular}{ccccc}
\hline
Sanders number &  $m_{\rm{V}}$ & $B-V$ & $P$(d) & Eccentricity \\
\hline
S0977&  10.04&  -0.10&         &                \\
S1434&  10.66&  0.12&          &            \\
S1066&  10.93&  0.11&          &            \\
S1267&  10.90&  0.22&   846    &     0.475      \\
S1284&  10.92&  0.24&   4.18       &     0.205      \\
S1263&  10.99&  0.21&          &            \\
S0968&  11.29&  0.12&          &            \\
S0975&  10.89&  0.43&   1221       &     0.088      \\
S1082&  11.21&  0.40&   1.068+1189 &     0.57       \\
S0752&  11.33&  0.29&   1003       &     0.317      \\
S1072&  11.28&  0.61&          &            \\
S1280&  12.22&  0.26&          &            \\
S0997&  12.11&  0.46&   4913       &     0.342      \\
S1195&  12.34&  0.42&   1154       &     0.066      \\
S0792&  12.00&  0.58&          &            \\
S0277&  12.28&  0.52&          &            \\
S2226&  12.50&  0.30&          &            \\
S1273&  12.27&  0.56&          &            \\
S0984&  12.19&  0.52&          &            \\
S1005&  12.68&  0.50&          &            \\
S0751&  12.69&  0.51&          &            \\
S1036&  12.84&  0.47&          &            \\
S0145&  12.90&  0.44&          &            \\
S2204&  12.88&  0.47&          &            \\
\hline
\end{tabular}
\end{center}
\end{table*}

\section{Summary and conclusions}
According to numerous observations of BSs in various stellar
systems, BSs in a given stellar system should be formed via
different formation mechanisms \citep{pio04}. Generally,
direct collisions are considered as the principal mechanism to
form BSs in a dense stellar environment \citep{fre04},
while mass transfer in a close binary system is regarded as the main
way to form BSs in a sparse environment \citep{mat91}.
Based on the scheme outlined long ago by McCrea (1964),
we have presented detailed modelling of PBs experiencing mass exchange
and simulated the situation of BSs in the old open cluster M67.

\par
The evolutionary model of BS formed via mass transfer is shown in
this work. As previous studies have been based on
simplified evolution schemes, detailed modelling of mass-transfer
binaries may have positive effects on population synthesis
and reveal more information on these binaries.
A binary of 1.4M$_\odot$+0.9M$_\odot$ was set as an example.
In this example, both the donor and the accretor follow
idiosyncratic evolutionary schemes that deviate completely from the
classical evolution scenario of single stars. By summing up the
total light of the two components in the example binary system
along their individual evolutionary tracks, we get a synthetic
evolutionary track of the example binary system in the CMD. According
to the synthetic track, the binary spends 1.228Gyrs in the region
where it is bluer than the MSTO of the 4.0Gyr isochrone until the secondary leaves
the main sequence. Thus the example binary lasts long enough in
the target region to be observed as a BS.

\par
A grid of primordial BS models for old stellar populations from
1.0Gyr to 6.0Gyr is also given. Using this grid, we can understand the BS
formed in this channel and simulate primordial BSs in real star
clusters in these ages. According to the grid, the formation process
of the PBs is continuous during the evolution of a cluster$-$like
M67. The PBs simulated by our calculation at a certain epoch
were produced via mass exchange, which happened before the time of
observations. At this epoch, a star whose mass is lower than the one
at the MSTO of the corresponding isochrone may evolve into a straggling
phase with respect to a single-star scenario by gaining material from
the donor, which brings the whole binary system into a peculiar
evolutionary scheme.

\par
Based on detailed modelling of primordial BSs, a Monte-Carlo
simulation of the old open cluster M67 was made. The simulation
defines an upper boundary of primordial BS in the CMD. Thus we can
define a primordial BS region between this upper limit and the given
isochrone. This region can better constrain the position of the BSs
formed via mass transfer in the CMD, which is helpful for
distinguishing the formation channels of the observed BSs. We would
like to argue that the BSs with short periods entering the
primordial BS region should have a mass-transfer origin. Our model
only predicts 4 primordial BSs (1/6 of the observed number count).
For the case of M67, there are two such instances (S1284 and the
close binary component of the triple system S1082). The observation
fairly agrees with our prediction for the distribution of orbital
inclinations of binary systems.

\par
Hurley et al. (2001) tested different combinations of these initial
parameter distributions in their binary population synthesis. The
maximum number of blue stragglers generated in their simulations is
8.2 per 5000 binaries at 4.2Gyr, while the minimum is 1.1 out of
5000. Obviously, these results cannot account for the number of
observed BSs in M67. Thus they conclude that cluster dynamics is
beneficial not only for explaining BSs with eccentric and/or wide
orbit but also for increasing the simulated BSs number. In their
dynamical simulations, 22 BSs are generated, but only one of them is
in a binary, while the predicted number of BSs reaches the maximum
at 3.653Gyr with seven of BSs in binaries. As a result, they thought
that the blue stragglers observed in M67 should be formed through
several processes. Recently, Hurley et al. (2005) presented a new
result of dynamical simulation using a direct N-body model and also
performed new BS population synthesis for M67. At 4Gyr, the new
dynamical model generates 20 blue stragglers with 9 of them in
binaries, while BS population syntheses produce 25 BSs with 6 in
binaries. From both studies, it is concluded that dynamical
processes destroy BSs as well as creating new ones. According to
their new results, Hurley et al. (2005) argue that both the
primordial binary population and the dynamical environment play
essential roles in generating BSs in open clusters. There is an
obvious difference between their results of population synthesis and
our simulations. The primary reason is that we are strictly limited
to PBs from case A. Instead, they have implemented many more BS
formation channels in their population synthesis, including Case B
and C mass transfer schemes and wind accretion. After all, inclusion
of tides and a distribution of eccentricities could also affect the
number of BSs produced via the Case A scheme.

\par
With the mechanism of primordial BS formation, the period
distribution of the PBs can be predicted. This is proved by a
comparison between the 2 observed BSs with short periods in M67 and
our theoretical results. This is a preliminary approach to this
observable property of primordial BSs, so more studies and finer grid
of binary models are needed.

\par
In fact, there are some primordial BS formation processes that are
not considered in the present work, such as the merger of short
period binaries and wind accretion in wide binaries, the latter
case can possibly account for the observed BSs in wide binaries.
Even within the primordial BS region, other mechanisms
are also needed. More observations, especially of binaries and
orbital parameters, are needed in order to understand the BS
population in M67.

\begin{acknowledgements} We would like to thank the National Science
Foundation of China (NSFC) for support through grants 10333060,
10573022, and 10521001. We are grateful to the referee for his/her
useful and inspirational comments that helped to improve this work.
\end{acknowledgements}

\clearpage

\end{document}